\documentclass[aps,twocolumn,notitlepage,superscriptaddress]{revtex4-2}

  \usepackage[utf8]{inputenc}
  \usepackage[english]{babel}
  \usepackage[T1]{fontenc}

  \usepackage{graphicx}
  \usepackage{amssymb,amsfonts,amsmath}
  \usepackage{braket}
  \usepackage{hyperref}
  \usepackage{color}
  \usepackage{soul}
  \usepackage{comment}
  \usepackage[normalem]{ulem}
  \usepackage{placeins}
  \usepackage{upgreek}

\begin{document}
  \title{Loss-independent time-delay estimation \\via Hong-Ou-Mandel photon bunching detection}

  \author{Julián Defant}%
  \email{defant.julian@gmail.com}
  \affiliation{Departamento de Física, Facultad de Ciencias Exactas y Naturales, UBA, Ciudad de Buenos Aires,
  Argentina}

   \author{Agustina G. Magnoni}
    \affiliation{Departamento de Física, Facultad de Ciencias Exactas y Naturales, UBA, Ciudad de Buenos Aires, Argentina}
  \affiliation{%
   Laboratorio de Óptica Cuántica, DEILAP, UNIDEF (CITEDEF-CONICET), Buenos Aires, Argentina}%

  \author{Miguel A. Larotonda}%
   \affiliation{Departamento de Física, Facultad de Ciencias Exactas y Naturales, UBA, Ciudad de Buenos Aires, Argentina}
  \affiliation{%
   Laboratorio de Óptica Cuántica, DEILAP, UNIDEF (CITEDEF-CONICET), Buenos Aires, Argentina}%

  \author{Laura T. Knoll}%
  \email{lknoll@citedef.gob.ar}
   \affiliation{Departamento de Física, Facultad de Ciencias Exactas y Naturales, UBA, Ciudad de Buenos Aires, Argentina}
  \affiliation{%
   Laboratorio de Óptica Cuántica, DEILAP, UNIDEF (CITEDEF-CONICET), Buenos Aires, Argentina}%

  \date{\today}

  \begin{abstract}

   In this work we study the application of two-photon correlated quantum states for high-precision time estimation. To achieve this, photon pairs were fed into a Hong-Ou-Mandel polarization interferometer to generate N00N entangled states with $N = 2$. Two- and four-detector schemes were considered, and the achievable time estimation precision of each was studied, with particular emphasis on the system's Fisher information in each case.
  Our results reveal that using a four-detector scheme,
  enabling the measurement of both coincidence and photon-bunching terms,
  achieves the maximum precision given by the quantum Fisher information, without the need of characterizing system losses. We also show that correlations between repetitions arise due to systematic errors, limiting the precision enhancement that can be obtained by increasing the number of repetitions. Taking these correlations into account, we show that the four-detector scheme outperforms the two-detector scheme in the precision of time-delay estimation. With the experimentally determined calibration uncertainties, the four-detector scheme improves the dispersion by a factor of 2 for the repetitions performed, reaching up to a factor of 4 as both schemes approach their systematic floors.
  \end{abstract}

  \maketitle

  \section{\label{sec:intro} Introduction}
  In the context of the rapid development of quantum technologies, the pursuit of physical measurements with ever-increasing levels of precision has become a fundamental challenge.
  In this scenario, photonic quantum metrology stands out as a key discipline, as it allows for the exploitation of quantum properties of light, such as entanglement and indistinguishability, to design measurement protocols that overcome the sensitivity limits imposed by classical physics \cite{giovannetti2004quantum,giovannetti2011advances,pirandola2018advances,dowling2015quantum,polino2020photonic}. At the heart of these high-precision techniques lies photon interferometry, where the Hong-Ou-Mandel (HOM) effect \cite{HOM} proves to be a fundamental resource for estimating optical delays \cite{chen2019hong,kuts2025precision,scott2020beyond,jordan2022quantum}.
  This effect occurs when two identical photons enter different input ports of a beamsplitter and exit together through the same port, creating a ``dip'' in the coincidence rate.
  HOM interference can be used to generate two-photon N00N states in a deterministic and straightforward manner. This family of states
  exploit photon indistinguishability to achieve phase measurements reaching the Heisenberg limit \cite{giovannetti2004quantum,giovannetti2011advances, mitchell2004super,knoll2019role}.

  In time-sensitive measurements, the HOM interferometer measures optical delays by observing the shift in the position of the dip. This principle is famously applied in quantum optical coherence tomography (QOCT) \cite{nasr2003demonstration}, enabling dispersion-canceled, micrometer-scale measurements applicable to biological samples \cite{taylor2016quantum, mauranyapin2017evanescent,cox2006optical}.

  Recent advances have focused on achieving greater information gain per photon by tuning the interferometer to the delay that contains the maximum information (defined by the Fisher information) instead of relying on detecting the shift in the HOM dip \cite{lyons2018attosecond, chen2019hong}.
  Alongside these phase-tuning techniques, state-of-the-art approaches have demonstrated that resolving the spectral profile of the interfering photons can further push the estimation near the quantum limit while significantly enhancing the dynamic range \cite{jordan2022quantum,lualdi2025fast,brooks2026optical}.

  However, despite these sophisticated measurement strategies, a major bottleneck in practical quantum metrology remains the presence of system losses \cite{scott2021noise, meskine2024approaching}. In standard interferometry, the loss of a single photon transforms the highly sensitive N00N state into an incoherent mixture of states occupying different number of photons N, thereby degrading the available information and severely bounding the attainable quantum advantage \cite{demkowicz2012elusive}. Standard two-detector schemes are particularly vulnerable to this issue, as losses introduce ambiguities between lost photon pairs and valid coincidences. Consequently, these schemes require precise characterization of the system's efficiency to accurately bound the estimation error.
  Furthermore, while theoretical models typically assume statistically independent measurements, implying that the estimation dispersion scales indefinitely as $1/\sqrt{m}$ with the number of repetitions $m$, real-world estimation is often limited by imperfect calibration. Uncertainties in systematic parameters (such as the interferometric visibility) act as nuisance parameters that introduce non-trivial correlations between successive experimental repetitions \cite{suzuki2020nuisance}. If left unaccounted for, these systematic errors impose a fundamental practical limit on the precision gained from accumulating data.

  In this work,
  we perform an experiment based on a two-photon polarization HOM interferometer and propose a conditional detection scheme whose estimator is insensitive to losses, similar to the one used in \cite{lualdi2025fast}. By detecting all photon pairs using four detectors, two per output mode of the HOM interferometer, we reach the quantum Fisher information (QFI) for perfect visibility conditions, which ultimately defines the attainable precision.
  We demonstrate, both theoretically and experimentally, that exploiting the full photon-pair detection statistics removes the need to calibrate the losses and the systematic floor this calibration introduces, while systematic correlations between repetitions impose a fundamental practical limit on the precision gained from increasing the number of measurements.

  \section{\label{sec:theory}Theoretical description}

  \subsection{Theory of parameter estimation}

  In quantum metrology, parameter estimation aims at inferring the value of an unknown physical parameter $s$ from the statistics of measurement outcomes obtained from a parameter-dependent quantum state. The parameter is encoded into a probe state through a physical process, resulting in a quantum state $\rho_s$. Information about $s$ is subsequently extracted by performing a measurement, described in general by a positive operator-valued measure (POVM) $\mathcal{M}={M_x}$, which gives rise to the conditional probability distribution
  \begin{equation}
  P(x|s,\vec{\xi})=\mathrm{Tr}\left[\rho_s M_x\right],
  \end{equation}
  where $\vec{\xi}$ denotes additional parameters characterizing the experimental system, which may include known parameters. The estimation problem therefore consists in constructing an estimator $\hat{s}$ from a set of measurement outcomes ${x_1,\ldots,x_m}$ such that its statistical properties allow the unknown parameter $s$ to be inferred with the highest possible precision.

  A standard approach for constructing an estimator is maximum-likelihood estimation (MLE). Given the observed outcomes and the probability distribution $P(x|s,\vec{\xi})$, the likelihood function $\mathcal{L}(s)$ quantifies the probability of obtaining the observed data for a given value of the parameter. For statistically independent measurements,
  %
  %
  the maximum-likelihood estimator $\hat{s}_{\mathrm{ML}}$ is obtained by maximizing $\mathcal{L}(s)$ (or equivalently, its log-likelihood function) with respect to $s$ \cite{lane1993maximum,knoll2025experimental}:
  %
  \begin{equation}\label{MLE}
  \frac{\partial \ln \mathcal{L}(s)}{\partial s}
  =
  \frac{\partial}{\partial s}
  \sum_{i=1}^{m}\ln P(x_i|s,\vec{\xi})
  =0.
  \end{equation}
  Under suitable regularity conditions, maximum-likelihood estimators are asymptotically unbiased and efficient, meaning that their variance approaches the fundamental statistical lower bound in the limit of a large number of independent measurements.

  For an unbiased estimator, the precision of the estimation is fundamentally constrained by the classical Cramér--Rao bound (CRB) \cite{rao1992information,fisher1922mathematical,toth2014quantum},
  \begin{equation}\label{CR}
  \mathrm{Var}(\hat{s})\geq \frac{1}{mF(s)},
  \end{equation}
  where $m$ is the number of independent repetitions and $F(s)$ is the Fisher information (FI) associated with the chosen measurement. The FI quantifies the sensitivity of the probability distribution to variations of the parameter and is defined as
  \begin{equation}\label{eq:FI}
  F(s)=
  \sum_x
  \frac{1}{P(x|s,\vec{\xi})}
  \left|
  \frac{\partial P(x|s,\vec{\xi})}{\partial s}
  \right|^2.
  \end{equation}
  Thus, the FI characterizes the amount of information about $s$ that can be extracted from a given measurement scheme. Importantly, it is a property not only of the probe state and the parameter-encoding process, but also of the measurement performed on the system.

  In quantum metrology, the measurement is not necessarily fixed a priori. Quantum mechanics allows one to optimize the extracted information over all physically allowed measurements. For a given parameter-dependent state $\rho_s$, the quantum Fisher information (QFI) is defined as the maximum classical FI attainable over all POVMs,
  \begin{equation}\label{eq:QFI}
  F_Q(s)=\max_{\mathcal{M}} F(s).
  \end{equation}
  The QFI therefore characterizes the maximum amount of information about $s$ that can, in principle, be extracted from the quantum state, independently of the particular measurement used to access it.


  Since $F(s)\leq F_Q(s)$,
  the quantum Cramér--Rao bound (QCRB) follows directly from the classical CRB \cite{pezze2014quantum,toth2014quantum,paris2009quantum, liu2020quantum},
  \begin{equation}\label{eq:QCRB}
  \mathrm{Var}(\hat{s})
  \geq
  \frac{1}{mF(s)}
  \geq
  \frac{1}{mF_Q(s)}.
  \end{equation}
  The QCRB thus establishes the ultimate precision limit imposed by quantum mechanics for the estimation of $s$, while the classical CRB quantifies the precision limit associated with a specific measurement strategy. Consequently, quantum-enhanced metrology seeks to optimize both the parameter-dependent probe state and the measurement strategy so as to maximize the QFI and, when possible, realize a measurement for which the classical FI reaches the QFI.

  \subsection{Polarization-based HOM interference and detection schemes}\label{sec:HOMpol}


   \begin{figure}[]
     \centering
      \includegraphics[width=0.35\textwidth]{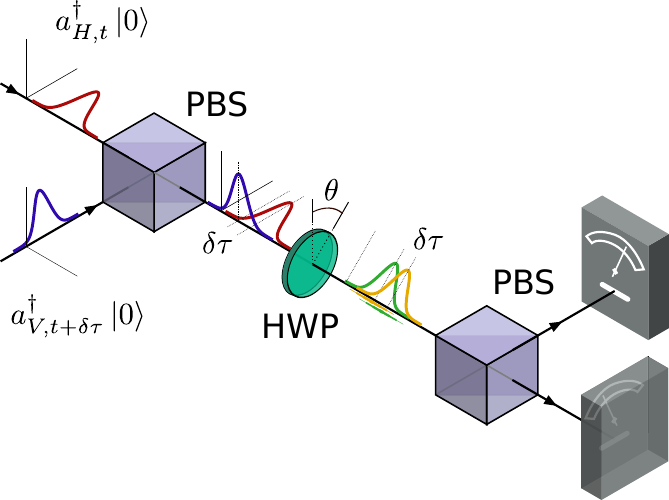}
      \caption{Schematic configuration of a standard polarization-based Hong-Ou-Mandel interferometer for time-delay estimation. }
      \label{PolScheme}
  \end{figure}

  We will consider a polarization-based HOM interferometer, where modes $H$ (horizontal) and $V$ (vertical) play the same role as path modes in the traditional HOM interferometer.
  Interference in the polarization degree of freedom is produced by an arrangement analogous to the BS in the traditional implementation, consisting of a polarizing beamsplitter (PBS) followed by a half-wave plate (HWP) and another PBS, as shown in Fig.~\ref{PolScheme}.
  Photons with orthogonal polarizations travel through orthogonal path modes and impinge on the input ports of the first PBS. Given that horizontally polarized photons are transmitted while vertically polarized ones are reflected, both photons exit the PBS through the same output port. In this way, the PBS allows for the combination of both beams in a single path. The photons are then directed to a half-wave plate 
  which applies the following transformation
  %
  \begin{equation}
       \begin{pmatrix}a^{\dagger}_H
   \\a^{\dagger}_V
  \end{pmatrix}\xrightarrow{\text{HWP}} -i \begin{pmatrix}
   \cos(2\theta)& \sin(2\theta) \\
   \sin(2\theta)& -\cos(2\theta) \\
  \end{pmatrix}  \begin{pmatrix}a^{\dagger}_H
   \\a^{\dagger}_V
  \end{pmatrix},
  \end{equation}
  where $a^{\dagger}_{H,V}$ are the creation operators for each polarization mode $H$ and $V$. If the photon pair is indistinguishable and the orientation of the HWP is $\theta = \pi/8$, the transformation reduces to
  \begin{equation}
      a^{\dagger}_Ha^{\dagger}_V\ket{0}\xrightarrow{\text{HWP}}\frac{1}{2}((a^{\dagger}_H)^2-(a^{\dagger}_V)^2)\ket{0},
  \end{equation}
  and the resulting state is a N00N state in the polarization degree of freedom, with $N=2$. 

  The HWP is followed by a second PBS, such that a pair of indistinguishable photons that enter the interferometric configuration exit the PBS through the same output port, resulting in a photon-bunching event. As the distinguishability of the photon pair increases, the probability of a coincidence event also rises, with each photon exiting the configuration through different output ports.

  The probe state of the system just after the first PBS, that is, the two-photon state before the HWP and thus \emph{before} interference, can be written as

  \begin{equation}\label{probe_state}
  \begin{split}
      \ket{\psi(\delta \tau)} = \int d\omega_1 d\omega_2 \;
          & f(\omega_1, \omega_2) \, e^{-i \frac{\delta \tau}{2} (\omega_1 - \omega_2)} \\
          & \times\, a^{\dagger}_H(\omega_1) \, a^{\dagger}_V(\omega_2) \ket{0},
  \end{split}
  \end{equation}
  where $\omega_i$ is the frequency of photon $i$, $f(\omega_1, \omega_2)$ is the joint spectral amplitude (JSA) of the state and $\delta \tau$ is the time delay between the arrival of the pair to the interferometer. As the unitary evolution associated with the interferometer does not depend on $\delta \tau$, the QFI for the time delay can be computed using directly the probe state described in Eq.~\eqref{probe_state} (see Appendix~\ref{app:A}) \cite{jordan2022quantum}, resulting in

  \begin{equation}\label{qfi_from_probe}
      F_Q =  \left( \Delta \omega_1 \right) ^2 + \left( \Delta \omega_2 \right) ^2 - 2 \, \mathrm{Cov}(\omega_1, \omega_2),
  \end{equation}
  where $\Delta \omega_i$ is the standard deviation of the frequency of photon $i$, and $\mathrm{Cov}(\omega_1, \omega_2)$ their covariance, both determined by the JSA of the state. For a separable JSA photon frequencies are uncorrelated and the covariance term vanishes. Considering a gaussian spectral amplitude, $\Delta \omega_i$ is the $1/e$ bandwidth of photon $i$, directly linked to the spectral filtering of the source. Assuming an identical bandwidth for both photons, $\Delta \omega_1 = \Delta \omega_2 \equiv \frac{2}{t_c}$, where $t_c = l_c / c$ is the coherence time and $l_c$ the full $1/e$ width of the HOM dip. Finally, in the absence of losses, the QFI for the time-delay estimation with a separable Gaussian JSA depends only on the coherence time of the pair source:

  \begin{equation}\label{QFItc}
      F_Q = \frac{8}{t_c^2}.
  \end{equation}
  Given a photon pair incident to the interferometric configuration, the measurable events, and therefore the classical Fisher information of the system, depend on the chosen detection scheme. The simplest scheme we will consider is based on two single-photon non-number-resolving detectors, one for each output path (we will refer to this configuration as $2Det$). This configuration allows the measurement of coincidences ($N_2$) and single-click events ($N_1$). The latter can be produced either by a pair of bunched photons arriving at the same detector or by a single photon whose twin was lost due to system efficiencies. In order to complete the measurement set, no-detection events ($N_0$), associated with lost photon pairs, must be taken into account. Then, the probability distributions of the possible events in a two-detector scheme are \cite{lyons2018attosecond}

  \begin{equation}
  \begin{aligned}
      &P_0(\delta \tau, \alpha, \gamma) = \gamma^2 \\
      &P_1(\delta \tau, \alpha, \gamma)= \frac{1}{2}(1-\gamma)^2 \left(\frac{1+3 \gamma}{1-\gamma} + \alpha e^{-4(\frac{\delta \tau}{t_c})^2}\right) \\
      &P_2(\delta \tau, \alpha, \gamma) = \frac{1}{2}(1-\gamma)^2 \left(1-\alpha e^{-4(\frac{\delta \tau}{t_c})^2}\right),
  \end{aligned}
  \label{eq:probas_atto}
  \end{equation}
  where
  $\alpha$ is the interferometric visibility and $\gamma$ is the loss probability per photon.
  Given these distributions, the Fisher information for the time delay $\delta \tau$ is given by
  \begin{equation}
  F_{2Det}(\delta\tau)=
  \frac{64}{t_c^2}
  \frac{\alpha^2(1-\gamma)^2(1+\gamma)
  \left(\dfrac{\delta\tau}{t_c}\right)^2
  e^{-8\left(\frac{\delta\tau}{t_c}\right)^2}
  }{
  \left[1-\alpha e^{-4\left(\frac{\delta\tau}{t_c}\right)^2}\right]
  \left[
  1+3\gamma+\alpha(1-\gamma)e^{-4\left(\frac{\delta\tau}{t_c}\right)^2}
  \right]
  }
  \end{equation}

  which clearly depends on the system losses $\gamma$.
  For the particular case of perfect visibility ($\alpha = 1$), we can calculate the maximum, given by the limit at $\delta\tau = 0$:
  \begin{equation}\label{maxF2det}
  \begin{split}
          \max F_{2Det}(\delta\tau)
          &= \lim_{\delta\tau\rightarrow0}F_{2Det}(\delta\tau)=\\
          &=\frac{8}{t_c^2}(1-\gamma)^2 = F_Q(1-\gamma)^2.
  \end{split}
  \end{equation}
  Therefore, the maximum Fisher information for this detection scheme  depends on the loss parameter $\gamma$, achieving the ultimate precision given by Eq.~\eqref{QFItc} only for the particular case of perfect visibility and no losses ($\gamma$=0).

  A more advanced detection scheme incorporates four detectors -- two for each interferometric output path (we will refer to this configuration as $4Det$). In this configuration, the light exiting each port of the interferometer is directed into a fiber beam splitter (FBS), which performs spatial multiplexing by routing the output photons toward the single-photon detectors. In this case, not only can coincidence events ($N_{11}$) be measured (which correspond to coincidences between detectors associated with different interferometer output paths), but also photon-bunching events ($N_{20}$ and $N_{02}$), which can be identified as coincidences between detectors from the same output path. Therefore, the complete measurement set of interest is restricted entirely to coincidence detections, significantly mitigating the ambiguities introduced by system losses and non-unit detection efficiencies. The probability distributions associated with the possible events in this scheme are \cite{ma2025unraveling}

  \begin{equation}\label{prob4det}
  \begin{aligned}
      P^{(1:1)}&(\delta \tau, \delta \omega, \theta) = \sin^4(2\theta) + \cos^4(2\theta) \\
      &- 2 \sin^2(2\theta) \cos^2(2\theta) \\
      &\times \exp\left[-4\left(\frac{\delta \tau}{t_c}\right)^2 - \frac{(t_c \delta \omega)^2}{16}\right], \\[0.3cm]
      P^{(2:0)}&(\delta \tau, \delta \omega, \theta) = P^{(0:2)}(\delta \tau, \delta \omega, \theta) \\
      &= \sin^2(2\theta) \cos^2(2\theta) \\
      &\times \left(1 + \exp\left[-4\left(\frac{\delta \tau}{t_c}\right)^2 - \frac{(t_c \delta \omega)^2}{16}\right]\right).
  \end{aligned}
  \end{equation}

  Here we introduce two parameters to account for any possible distinguishability between the photons in degrees of freedom other than the arrival times: $\delta\omega = \omega_2 - \omega_1 = \frac{2\pi c}{\lambda_0^2}\,\delta\lambda$ is the angular frequency difference of the photon pair, and $\theta$ is the orientation of the HWP defining the photon polarization. We define the interferometric visibility $\alpha = e^{-t_c^2\delta\omega^2/16}$ to match the definition of the previous scheme, which in practice also absorbs any other source of distinguishability; we therefore treat $\alpha$ as the calibrated parameter. In this case, the Fisher information for the $\delta \tau$ parameter can be computed numerically.

  It is very convenient at this point to introduce the dimensionless parameter $s = \frac{\delta \tau}{t_c}$. Assuming that the frequency difference $\delta \omega$ is fixed by the photon source, the distinguishability of the photon pair is completely encoded in $s$. The dependency of the probability distributions of Eq.~\eqref{prob4det} with $s$ is shown in Fig.~\ref{simulations}(a), where we assumed $\delta \omega=0$ and $\theta=\pi/8$ such that the indistinguishability condition ($s = 0$) leads to the HOM effect.
  For these conditions we can compute the Fisher information for the $4Det$ system using Eq.~\eqref{eq:FI}:

  \begin{equation}
      F_{4Det}(\delta \tau,\delta \omega=0, \theta=\pi/8) = \frac{64 \, \delta \tau^2}{t_c^4} \frac{e^{-8 \, \delta \tau^2 / t_c^2}}{1 - e^{-8 \, \delta \tau^2 / t_c^2}}.
  \end{equation}

  As in Eq.~\eqref{maxF2det}, its maximum value can be calculated by the limit $\delta \tau \rightarrow 0$, which results in

  \begin{equation}\label{maxF4det}
      \max F_{4Det}(\delta\tau)= \frac{8}{t_c^2}= F_Q,
  \end{equation}
  proving that our scheme achieves the ultimate precision under the  perfect visibility condition without the need of characterizing losses.
  For a more general and realistic case, with the probability distributions described for each detection scheme, we can construct an estimator for the parameter $s$ and establish the attainable precision in each case.

  \subsection{Maximum-likelihood estimation and dispersion floor}\label{sec:sims}

  \begin{figure*}[]
      \centering
      \includegraphics[width=0.8\textwidth]{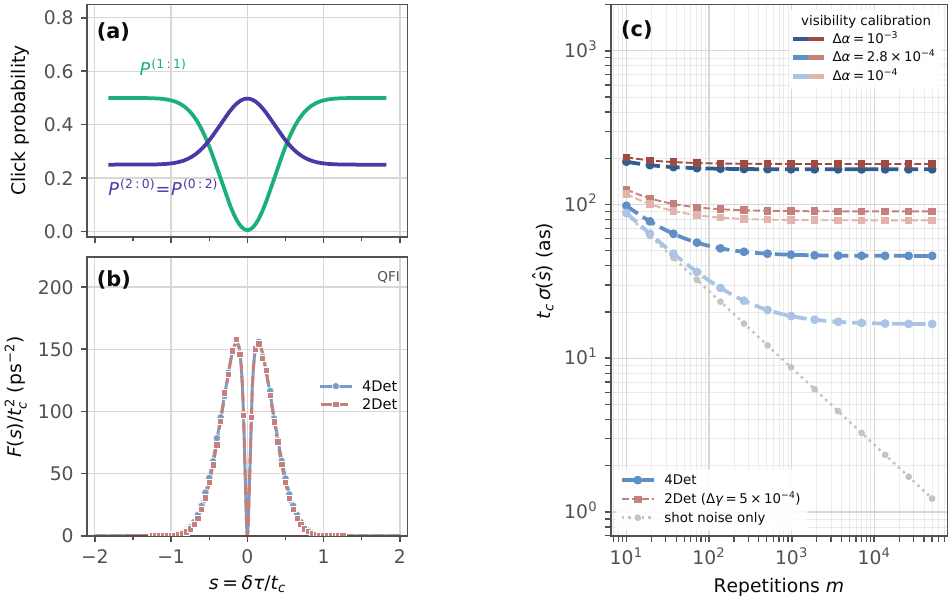}
      \caption{(a) Theoretical probability distributions for coincidence ($P^{(1:1)}$) and bunching ($P^{(2:0)}$, $P^{(0:2)}$) events as functions of $s = \frac{\delta \tau}{t_c}$. (b) Simulated Fisher Information for the four-detector ($4Det$) and two-detector ($2Det$) scheme according to its theoretical description. (c) Simulated dispersion of the estimation of $s$ considering correlation between repetitions due to the presence of the systematic parameter $\alpha$ for different values of its uncertainty. In blue, the $4Det$ scheme simulations for each calibration uncertainty considered; in red/brown, simulations corresponding to $2Det$ system, where the loss uncertainty is also taken into account. }
      \label{simulations}
  \end{figure*}

  For the two-detector scheme, the MLE is given by \cite{lyons2018attosecond}

  \begin{equation}
      \hat{s}_{2Det} = \pm \frac{1}{2}\sqrt{\ln \left[ \frac{\alpha(N_1 + N_2)}{N_1 - N_2(\frac{1+3 \gamma}{1 - \gamma})}\right]}.
  \end{equation}
  Here, $N_1$ ($N_2$) represents the number of times the experimental outcome was a single-click (coincidence) event. The four-detector scheme MLE $\hat{s}_{4Det}$ can be computed numerically solving Eq.~\eqref{MLE}, and depends on the number of times each possible result was measured ($N_{20}$ and $N_{02}$ for bunching events and $N_{11}$ for coincidence events).

  To simulate the experiment, the outcome of a single repetition is modeled as a draw from a multinomial distribution with three possible outcomes, whose probabilities depend on $s$ and differ between detection schemes. The meaning of the number of trials also differs: in the two-detector scheme it is the number of photon pairs \emph{entering} the interferometer, whereas in the four-detector scheme it is the number of \emph{detected} pairs. For a given $s$, each simulated repetition returns three counts, one per outcome. These, combined with fixed values of the systematic parameters, yield an estimate of $s$ through the MLE. To reconstruct the Fisher information we compute the variance of the estimator over $m$ repetitions and take Eq.~\eqref{CR} at equality, which yields $F(s)$. Repeating this over a range of $s$ recovers the dependence of the Fisher information on the time delay.

  Reproduction of the experimental conditions requires assigning values to the systematic parameters. The visibility $\alpha$ represents the quality of the interference and depends not only on the distinguishability of the pair, that is, on $\delta\lambda$, but primarily on the alignment of the interferometer. Both effects reduce the visibility, so in the four-detector scheme we absorb them into a single effective $\alpha$.

  Figure~\ref{simulations}(b) shows the simulated Fisher information for both schemes, together with the theoretical predictions calculated using Eq.~\eqref{eq:FI} and the corresponding probability distributions. The parameters were fixed from the experimental conditions: the interference visibility was obtained from the coincidence dip ($\alpha = 0.99$, corresponding to $\delta\lambda \approx 0.7$ nm), the coherence time from the width of the HOM dip ($t_c = 0.2$ ps), and the photon loss rate of the system is $\gamma = 0.88$. In both cases the variance of the estimator saturates the Cramér–Rao bound, so each scheme attains the maximum precision allowed by its own statistics. Note that the Fisher information vanishes at the indistinguishability condition $s=0$, and has peaks at two symmetric values, where the coincidence probability distribution has maximum slope. Therefore, all estimates from here on will be performed at these points, where the estimation sensitivity of the system is maximal.


  It is instructive to compare the Fisher information attained by each scheme with the QFI of the system. Replacing the value of the measured coherence time in Eq.~\eqref{QFItc}, the Fisher information is bounded by $F_Q = 200$~ps$^{-2}$. Under its respective natural units (information per incident pair for the two-detector scheme and per detected pair for the four-detector scheme) the four-detector scheme reaches $78\%$ of this bound, whereas the two-detector scheme attains only $1\%$. The $2Det$ Fisher information reaches the same value as $4Det$ after normalizing by the detection probability $\eta^2 = (1-\gamma)^2$, that is, when expressed per detected pair, as Fig.~\ref{simulations}(b) shows. In the ideal-visibility limit ($\alpha = 1$, $\delta\lambda = 0$) and for $\delta\tau \to 0$, both schemes saturate the quantum Cramér–Rao bound per detected pair, extracting the maximum information available in the system.
  In a realistic scenario, near the maximum of the Fisher information both schemes extract essentially the same information per detected pair; losses reduce the two-detector information only away from this optimum.
  At the cost of a reduced detection rate, our scheme requires no characterization of the system losses to perform an estimation, while the two-detector scheme needs an external calibration of the loss probability $\gamma$. As we will show, this requirement degrades the maximum precision attainable for the estimation of the time delay.

  The value of $s$ estimated by the MLE depends not only on the collected data but also on the systematic parameters, which we arrange in a parameter vector $\vec{\xi}$. These are  typically determined once, with a finite uncertainty, and used afterwards in every repetition of the estimation, introducing correlations between repetitions. As a result, accumulating repetitions of the experiment to reduce the variance of the estimation no longer follows the usual $1/m$ scaling with the number of repetitions $m$, since the repetitions are no longer statistically independent. The correlation introduced by each systematic parameter must therefore be accounted
  for through the off-diagonal terms of the covariance matrix, that involve the derivatives $\partial\hat{s}/\partial\xi_k$. Rather than differentiating each estimator, we evaluate the sensitivity of the MLE numerically, which treats both detection schemes on the same footing: keeping the measured counts fixed, each $\xi_k$ is drawn from a normal distribution with mean equal to its calibrated value and standard deviation equal to its calibration uncertainty $\Delta\xi_k$, and $\mathrm{Var}(\hat{s})_{\xi_k}$ is the variance of the estimator over these draws. To first order,
  \begin{equation}
      \left( \frac{\partial \hat{s}}{\partial \xi_k} \right)^2 \left( \Delta \xi_k \right)^2 \simeq \mathrm{Var}(\hat{s})_{\xi_k}.
  \end{equation}
  Since the systematic parameters are calibrated independently, no cross terms appear, and the total dispersion of the estimate after $m$ repetitions is
  \begin{equation}
      \sigma(\hat{s}) = \sqrt{\frac{\mathrm{Var}(\hat{s})_{\mathrm{rep}}}{m} + \sum_k \mathrm{Var}(\hat{s})_{\xi_k}},
  \end{equation}
  where $\mathrm{Var}(\hat{s})_{\mathrm{rep}}$ is the variance of the estimator over repetitions. The second term does not decrease with $m$ and sets the floor of the dispersion.

  Figure~\ref{simulations}(c) shows the simulated dispersion of s for $4Det$ scheme both with genuine independent repetitions and including the correlations induced by the systematic parameters $\vec{\xi}$.  The gray dashed curve depicts the expected behavior when no correlations are present, hence the dispersion scales as the shot noise. The blue curves/dots take into account the correlations, in which only $\alpha$ is considered since operating at $\theta = \pi/8$ makes the estimator insensitive to the angle setting at first order. Different values of uncertainty for $\alpha$ are considered.

  The smaller $\Delta \alpha$, the more repetitions are needed before the correlation term dominates; once it does, the dispersion no longer decreases with $m$, and its value is set by the precision with which $\alpha$ was determined. This sets the minimum dispersion achievable by accumulating repetitions: once this floor is reached, the dispersion becomes independent of $m$, and its value is set directly by the precision with which the systematic parameter was determined. For instance, $\Delta \alpha = 10^{-4}~$ sets a floor of $17~$as that can be reached with $m \approx 300$ repetitions.
  Furthermore, the $2Det$ estimator not only depends on the interferometric visibility $\alpha$ but also on the system losses $\gamma$; therefore, another correlation term must be included in order to compute the dispersion of the estimation, and the dispersion floor will increase proportionally to the calibration uncertainty of the associated systematic parameter. The simulated evolution of the dispersion in the $2Det$ system is shown in red in Fig.~\ref{simulations}(c) considering $\Delta \gamma = 5 \times 10^{-4}$. Given a fixed uncertainty on $\alpha$, the minimum dispersion achievable by the two-detector scheme is always higher than the corresponding four-detector dispersion due to correlations induced by the presence of $\gamma$ in the $2Det$ probability distributions. For instance, the dispersion floor in the $\Delta \alpha = 10^{-4}$ scenario rises to $79~$as, more than four times the respective floor achievable in the four-detector configuration.

  In summary, when repetitions of the estimation process are correlated by the presence of systematic parameters in the probability distributions of the experiment, the achievable precision is determined not only by the number of repetitions, but also by the uncertainty with which the systematic parameters were measured.

  In the following section we put these results to test and we experimentally obtain practical uncertainty limits for the two detection schemes on the same setup.

  \section{\label{sec:exp} Experimental implementation}

  \begin{figure*}[]
      \centering
      \includegraphics[width=0.8\textwidth]{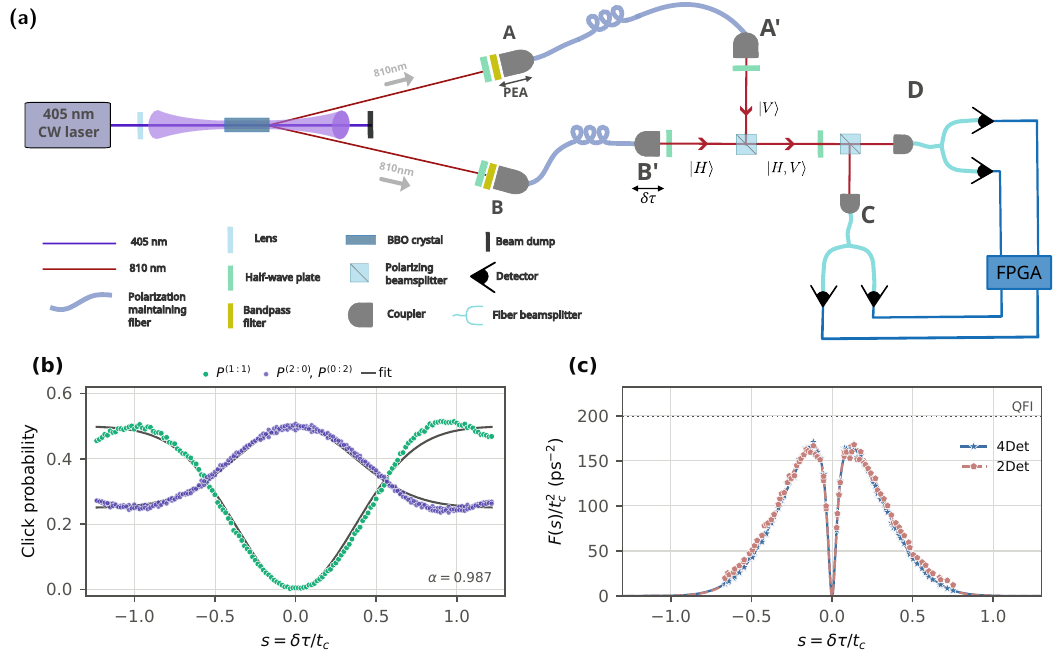}
      \caption{(a) Schematic setup of the Hong-Ou-Mandel polarization interferometer. Photon pairs at 810 nm are generated via SPDC using a CW 405 nm pump. After spectral filtering, photons are coupled into PMFs and redirected to the interferometer with orthogonal polarizations, where a PBS overlaps their paths. To scan the interference pattern, one of the fiber couplers is mounted on a motorized micrometer stage, enabling control over the relative delay between the photon pair. Interference is produced by an arrangement of a HWP followed by a second PBS, which separates the polarization components into different paths leading to the detection scheme. Each path is coupled to a FBS that divides the incident light spatially and leads the resulting photons into the detectors. The coincidences are computed and registered by an FPGA. (b) Probability measurements for coincidence events (green) and photon-bunching events (purple) with its respective theoretical prediction (black) for different time delays, which brings up the HOM dip. (c) Reconstruction of the Fisher information simulating multinomial repetitions based on measurements in (b) for the four-detector (blue) and two-detector (red) scheme, with its respective theoretical predictions.}
      \label{HOMscheme}
  \end{figure*}

  The experimental setup shown in Fig.~\ref{HOMscheme} comprises three main blocks: the correlated pair source, where photon pairs are generated by spontaneous parametric down-conversion (SPDC); the interferometric configuration, where the two photons in each pair interfere through their polarization degrees of freedom; and the detection scheme.
  The correlated pair source consists of a CW 405 nm pump laser set to a power of 120 mW and a type-I BBO crystal generating photon pairs that exit the crystal with vertical polarization, following different spatial paths, as depicted in Fig.~\ref{HOMscheme}. Before being collected and coupled into a polarization-maintaining fiber (PMF), each photon passes through a  10 nm bandwidth (FWHM) interference filter centered at 810 nm and a HWP that aligns the photon's polarization with the fiber axis.
   Photon pairs are routed to the interferometric configuration via the PMF and coupled out of the fiber into free space. One of the couplers is mounted on a motorized linear micropositioner stage, which controls the optical path difference between the photons and, therefore, the time delay $\delta\tau$. Once in free space, a second HWP is used to set the polarizations of the photon pair orthogonal to each other, as shown in Fig.~\ref{HOMscheme}. The first PBS of the interferometric configuration recombines the optical paths of the pair; the resulting polarization state $a_H^{\dagger}a_V^{\dagger} \ket{0}$ goes through an arrangement of a HWP set at $\theta=22.5$° and a second PBS, giving rise to a N00N state, as discussed in Sec.~\ref{sec:HOMpol}.

  Each output of the polarizing beamsplitter is coupled to one input of a fiber beamsplitter (FBS), whose two output ports are connected to separate single-photon avalanche photodiodes (APDs), yielding four detection channels in total. This is the $4Det$ scheme. It is important to notice that the information of the $2Det$ scheme can be fully recovered from these measurements. Finally, coincidences between every pair of detectors are computed and registered by an in-house built FPGA counter. Coincidences between detectors corresponding to the same output port of the interferometer are classified as photon-bunching events, while coincidences between detectors associated to different output ports are considered coincidence events.

  In Fig.~\ref{HOMscheme}(b) we show experimental measurements of the click probabilities for coincidence and photon-bunching events, recovering the usual dip in coincidences for indistinguishable photons ($s=0$) and a correlated peak in bunched pairs at the same point. The discrepancy between the measurements and the theoretical probability distributions in the wings of the dip is an artifact of the filtering process: the transfer function of the interferometric filters modulates the dip profile, resulting in a higher coincidence probability toward the edges. For larger values of $s$, the coincidence probability stabilizes at $\frac{1}{2}$, as expected \cite{ma2025unraveling}.

  The visibility $\alpha$ was estimated from click probabilities measured at the dip centre ($s=0$), where it follows directly from the ratio between coincidences and bunching events recorded in the same acquisition windows, without the need for any external reference or correction besides the accidental coincidences subtraction. In contrast, the loss parameter $\gamma$ must be obtained from click probabilities measured far from the dip, where the model assumes that the photons no longer interfere. A residual interference in these regions, which differs between both sides of the dip, biases the estimation of $\gamma$ and limits its accuracy. As a result, the calibration uncertainty of $\gamma$ ($\Delta\gamma = 1.6\times10^{-3}$) is considerably larger than that of $\alpha$ ($\Delta\alpha = 2.8\times10^{-4}$).

  In order to build the Fisher information of the system, the dip must be measured several times under the same experimental conditions, so that the procedure described in Sec.~\ref{sec:sims} can be implemented using measurements instead of simulated data. In our setup, visibility fluctuations caused by alignment instability limited the  repeatability of such measurement. Instead, the measurements in Fig.~\ref{HOMscheme}(b) were used as a seed to simulate repetitions via multinomial resampling. The Fisher information reconstruction using this method is shown in Fig.~\ref{HOMscheme}(c) for both detection schemes. The reconstructed Fisher information reaches about $80\%$ of the QFI, limited by the residual distinguishability ($\alpha = 0,987$).

  \begin{figure*}[]
     \centering
      \includegraphics[width=0.8\textwidth]{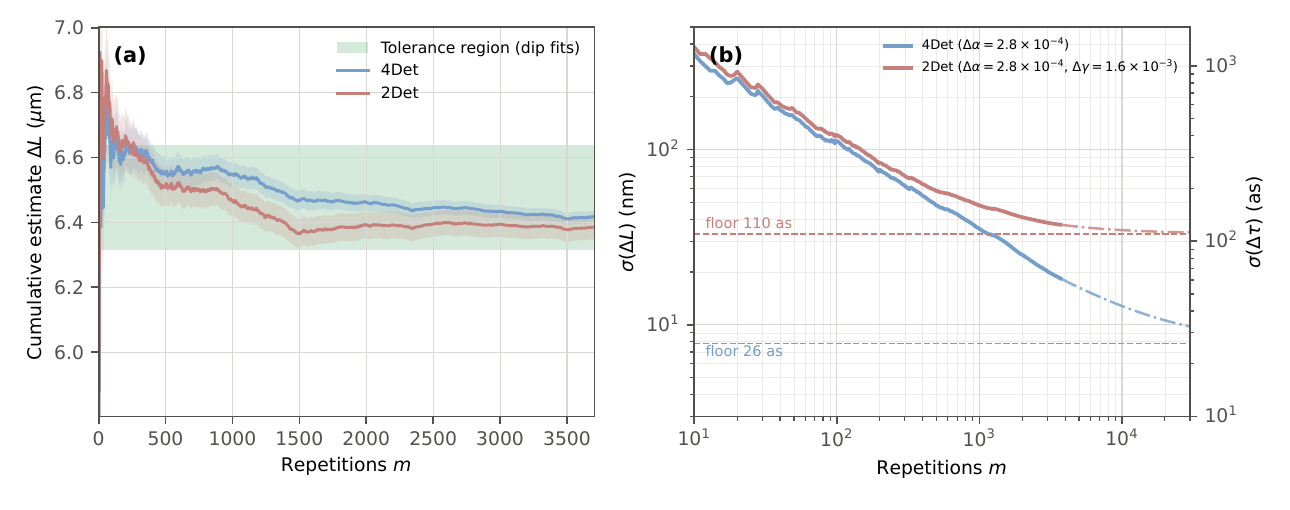}
      \caption{(a) Cumulative mean of the estimation of the path difference introduced by the piezoelectric actuator over repetitions for the two-detector scheme (red) and four-detector scheme (blue). (b) Cumulative dispersion of the path difference estimation over repetitions for the two-detector scheme (red) and four-detector scheme (blue). Horizontal dashed lines represent the dispersion floor set by the considered calibration uncertainty of the systematic parameters for each configuration.}
      \label{Results}
  \end{figure*}

  To introduce a controlled and reproducible path difference, a piezoelectric actuator (PEA) was mounted on coupler A (Fig.~\ref{HOMscheme}(a)). Switching the actuator between two fixed voltages displaces the coupler by a reproducible amount of a few micrometers (nominally 7 $\upmu$m), which shifts the photon delay between two operating points, $s_L$ and $s_H$, on the same side of the HOM dip. Both operating points are chosen in the neighborhood of the maximum of the Fisher information so that each delay is estimated with near-optimal precision. At each point we record the click statistics and obtain the corresponding delay with the maximum-likelihood estimator of each detection scheme (Sec.~\ref{sec:sims}): ($N_{11}$, $N_{20}$, $N_{02}$) for the four-detector scheme and ($N_1$, $N_2$) for the two-detector scheme. Both estimates are computed from the same detection events. The path difference then follows as $\Delta L = l_c \, (\hat{s}_H - \hat{s}_L)$.

   To cancel slow drifts of the interferometer, the measurement alternates between the two operating points on a time scale much shorter than that of the drift. Any drift is then common to both points of each pair and cancels in their difference, while the dip position and visibility are periodically tracked to keep the operating points fixed. Each pair of measurements constitutes one repetition. Figure~\ref{Results}(a) shows the cumulative mean of the estimated path difference for each detection scheme as a function of the number of repetitions, and Fig.~\ref{Results}(b) the corresponding cumulative dispersion, considering the correlation terms associated with the visibility parameter $\alpha$ for both schemes and the loss $\gamma$ for the two-detector configuration; horizontal dashed lines represent the dispersion floor set by the calibration uncertainty of these parameters.

\begin{figure}[]
     \centering
      \includegraphics[width=0.4\textwidth]{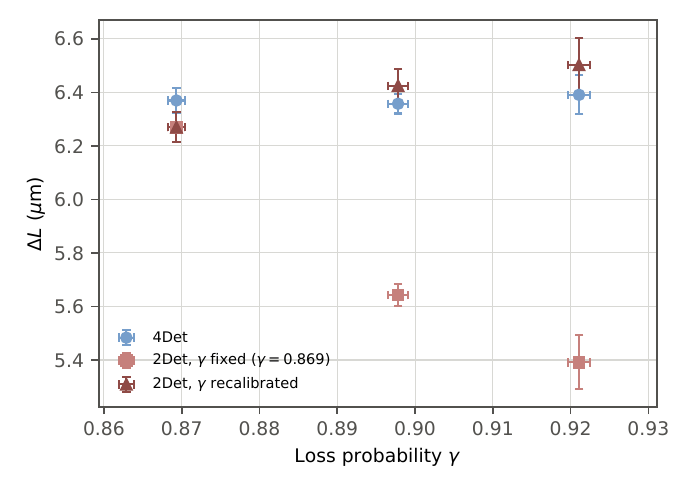}
      \caption{Estimated piezoelectric displacement $\Delta L$ versus the measured loss probability $\gamma$ for the $4Det$ scheme (blue circles), the $2Det$ scheme with $\gamma$ fixed at its first-configuration value (light red squares) and the $2Det$ scheme recalibrating $\gamma$ in each configuration (dark red triangles). Vertical error bars include the statistical error and the calibration systematics; horizontal error bars show the uncertainty in $\gamma$.}
      \label{Loss}
  \end{figure}

   As shown in Fig.~\ref{Results}(a), after $3.7\times10^{3}$ repetitions both detection schemes converge to estimates that do not differ significantly, $\Delta L = (6.418 \pm 0.018)~\upmu$m for the four-detector scheme and $(6.386 \pm 0.037)~\upmu$m for the two-detector scheme. Furthermore, after 400 repetitions, both estimates evolve inside the tolerance region determined by fitting the dip at each piezoelectric operating point. Remarkably, Fig.~\ref{Results}(b) shows that the dispersions of the two schemes start to differ after a few hundred repetitions, when the correlated calibration terms become relevant. At $3.7\times10^{3}$ repetitions, the two-detector dispersion is close to its floor of 33 nm which corresponds to a time delay of 110 as, set by the calibration uncertainties of $\alpha$ and $\gamma$. The four-detector dispersion, in contrast, is still statistically limited and well above its floor of 8 nm or 26 as, which is set only by the experimental determination of $\alpha$. With the current number of repetitions the dispersions already differ by a factor of about 2, and this factor can reach about 4 as both schemes approach their technical limits.
  To show that the accuracy of the two-detector estimate depends on the loss calibration, the piezoelectric displacement was estimated for three loss conditions, obtained by placing glass slides between the first PBS and the common HWP ($\gamma = 0.869$, $0.898$ and $0.921$) without realigning the interferometer, such that the visibility remained unchanged. For each condition, the same repetitions were analyzed for the $4Det$ scheme; for the $2Det$ scheme using the value of $\gamma$ measured in the first configuration, and finally for the $2Det$ scheme recalibrating $\gamma$ in each configuration.
  These results are shown in Fig.~\ref{Loss}.

  When considering a loss-varying system, the $2Det$ scheme delivers estimations that are significantly different for the same displacement when fixing the value of $\gamma$; only after a new measurement of the systematic parameter $\gamma$ do the estimates agree, within uncertainties, with those of the $4Det$ scheme. Remarkably, the changes in the loss parameter do not affect the four-detector scheme estimation, yielding estimates that agree for the three conditions without the need of an extra determination.

  \section{Concluding remarks}

  We have developed a theoretical and experimental study for the estimation of a time delay based on a HOM interferometer comparing two detection schemes, each based on two and four detectors, respectively. We characterized the achievable precision, quantified by the Fisher information for each case. Both schemes extract the same Fisher information per detected pair; the two-detector estimator, however, requires the loss probability as an external calibration parameter, while the four-detector estimator is independent of it.

  Although increasing the number of repetitions ($m$) can improve precision in the ideal independent-sample regime, our work reveals the presence of correlations between repetitions caused by systematic effects. Consequently, the expected $1/\sqrt{m}$ dispersion scaling cannot be extrapolated indefinitely. By considering these correlations, we show that the four-detector scheme achieves higher precision than the two-detector scheme, which is strongly limited by losses. By introducing a controlled and repeatable path difference
   with a piezoelectric actuator, we obtained a $2\times$ precision improvement with the four-detector scheme compared to the two-detector scheme which arises entirely from the loss-calibration floor, since both schemes reach the same statistical dispersion. Increasing the number of repetitions would allow us to achieve a $4\times$ difference, bounded by the correlated calibration parameters. When the losses were changed, the two-detector estimate shifted by up to 0.9 $\upmu$m unless $\gamma$ was recalibrated, while the four-detector estimate remained within 0.03 $\upmu$m.

The four-detector scheme does not simply improve photon detection: it removes the need to independently calibrate the system loss probability, while retaining the information available per detected photon pair. Our results provide both a practical route to loss-independent time-delay estimation and a framework for evaluating the limitations imposed by systematic parameter uncertainties in repeated quantum measurements.
These findings can significantly contribute to the design of robust and high-precision time-delay estimation schemes in quantum metrology.

  \begin{acknowledgments}
  All the authors acknowledge financial support from Consejo Nacional de Investigaciones Científicas y Técnicas (AR) and Agencia Nacional de Promoción Científica y Tecnológica (PICT-0201).
  The authors would like to thank Dario Rodrigues for stimulating and helpful discussions.
  \end{acknowledgments}

  \appendix
  \section{Quantum Fisher Information}
  \label{app:A}
  For the particular case of a pure parametrized state $\ket{\psi}=\ket{\psi(\delta \tau)}$ as the one described by Eq.~\eqref{probe_state}, the QFI can be easily calculated as \cite{liu2020quantum}
  \begin{equation}\label{qfi_pure}
     F_Q = 4\left(\braket{\partial_{\delta \tau}\psi|\partial_{\delta \tau}\psi}-|\braket{\partial_{\delta \tau}\psi|\psi}|^2 \right),
  \end{equation}
  where $\ket{\partial_{\delta \tau}\psi}$ is the partial derivative of $\ket{\psi}$ with respect to $\delta \tau$.
  Assuming a Gaussian JSA, our probe state can be written as:
  \begin{widetext}
  \begin{align}
  \ket{\psi(\delta \tau)} = &\int d\omega_1 d\omega_2 \; \left(\frac{1}{2\pi \Delta^2}\right)^{1/2}
          \exp\left[\left(\frac{-1}{4 \Delta^2}\right) (\omega_1^2 + \omega_2^2)+i(\omega_1-\omega_2)\frac{\delta \tau}{2}+i(\omega_1+\omega_2)\frac{t_1+t_2}{2}\right]
          a^{\dagger}_H(\omega_1) \, a^{\dagger}_V(\omega_2) \ket{0},
    \label{probe_gauss}
  \end{align}
  \end{widetext}
  where $t_1$ and $t_2$ are the arrival times of each photon, whose mean only contributes a phase independent of $\delta \tau$, and $\Delta=2/t_c$ the frequency bandwidth. Then, it is easy to see that
  \begin{equation}
      \braket{\partial_{\delta \tau}\psi|\partial_{\delta \tau}\psi}= \frac{1}{4}\braket{(\omega_1-\omega_2)^2}=\frac{1}{4}\left(\braket{\omega_1^2}+\braket{\omega_2^2}-2\braket{\omega_1\omega_2}\right),
  \end{equation}
  and
  \begin{equation}
      |\braket{\partial_{\delta \tau}\psi|\psi}|^2= \frac{1}{4}(\braket{\omega_1}-\braket{\omega_2})^2=\frac{1}{4}\left(\braket{\omega_1}^2+\braket{\omega_2}^2-2\braket{\omega_1}\braket{\omega_2}\right).
  \end{equation}
  Combining these results into Eq.~\eqref{qfi_pure} we finally obtain the expression for the QFI given in Eq.~\eqref{qfi_from_probe}.

  \bibliographystyle{ieeetr}
  \bibliography{biblio}








  \end{document}